# The Role of Generative AI in Strengthening Secure Software Coding Practices: A Systematic Perspective

Hathal S. Alwageed [†]
College of Computer and Information Sciences, Jouf University, Sakaka 42421, Saudi Arabia
hswageed@ju.edu.sa

Rafiq Ahmad Khan
Software Engineering Research Group, Department of Computer Science and IT, University of Malakand, Pakistan
rafiqahmadk@gmail.com

## ABSTRACT

As software security threats continue to evolve, the demand for innovative ways of securing coding has tremendously grown. The integration of Generative AI (GenAI) into software development holds significant potential for improving secure coding practices. This paper aims at systematically studying the impact of GenAI in enhancing secure coding practices from improving software security, setting forth its potential benefits, challenges, and implications. To outline the contribution of AI driven code generation tools, we analyze via a structured review of recent literature, application to the industry, and empirical studies on how these tools help to mitigate security risks, comply with the secure coding standards, and make software development efficient. We hope that our findings will benefit researchers, software engineers and cybersecurity professionals alike in integrating GenAI into a secure development workflow without losing the advantages GenAI provides. Finally, the state of the art advances and future directions of AI assisted in secure software engineering discussed in this study can contribute to the ongoing discourse on AI assisted in secure software engineering.

## 1 Introduction

Nowadays, reliable software systems are one of the critical needs in the functioning of all global sectors such as healthcare, banking, finance, transportation, defense, etc. However, ever growing dependence on interconnected systems have made us more susceptible to cybersecurity threats. Companies need to emphasize secure coding practices so that their systems are resilient to attack by malicious actors [1, 2]. Traditional cyber defense has lost its effectiveness, since conventional cyber threats have become more advanced and necessitate more competent protective measures [3, 4]. This requires new approaches that exploit the capabilities of Generative Artificial Intelligence (GenAI) to improve cybersecurity [5].

Serious problems can arise as a result of exploiting software security vulnerabilities [6]. Therefore, it is essential to provide new methods, techniques, and tools to assist teams in creating secure code within the first phases of software development to minimize costs and increase time to market. Boost, SVM radial, decision tree, random forest, and SVM linear are some machine learning methods used to construct trustworthiness models by Medeiros et al. [9]. This method evaluates the trustworthiness level or category of code units (such as files, classes, or functions/methods) using evidence from software security rather than to forecast or identify vulnerabilities. Graph neural networks are the cutting edge of artificial neural network research and development [10]. When analyzing software vulnerabilities, the Graph Neural Networks method performs well [10]. To identify software vulnerabilities, supervised and deep learning methods are currently popular. These methods have a high detection performance—up to 95% of the F1 score—and can successfully identify vulnerabilities, including cross-site scripting, SQL injection, and buffer overflow [11]. Applying various deep learning techniques to the vulnerability detection problem, Li et al. [12] compared and contrasted CNN, RNN, and Multilayer Perceptron (MLP) on both the SARD and NVD datasets. A range of multiclass classification algorithms were employed to estimate vulnerability vectors, including Naïve Bayes, decision tree, k-nearest neighbors, multilayer perceptron, and random forest. For feature extraction, natural language processing techniques like a bag of words, term frequency-inverse document frequency, and n-gram were also employed [13]. A multi-probability and difficult-to-predict problem can be effectively tackled by combining several approaches and classification algorithms, as demonstrated by their results.

Software security is an ever-present issue in the field [7]14]. Many different approaches have been suggested for finding and fixing vulnerabilities [8]. However, there has been an uptick in the number of security holes reported to the Common Vulnerabilities and Exposures (CVE) database [15]. Human experience, technology for project management, and code analysis all contribute to a weak defense [16, 17]. It suggests using a mix of machine learning and more conventional approaches to vulnerability discovery [17]. Software engineering vulnerabilities have been the subject of several solutions within the last twenty years [3].

It is clear from the preceding that safeguarding software applications during their development phases is insufficient. Discovering safer AI methods and safeguards for software programmers is an urgent matter [9]. Integrating AI frameworks with maturity models is pivotal in aligning AI adoption with organizational readiness and strategic growth. AI frameworks, encompassing tools, methodologies, and technologies for developing and deploying AI solutions, provide the technical foundation for automating processes, enhancing decision-making, and improving predictive capabilities. Professional tools evaluate how well organizations handle multiple dimensions like strategy and technology through defined stages of development. Combining AI frameworks with maturity models helps organizations move beyond user-dependent AI use and creates alignment that produces optimal results from technology adoption despite minimized risks. For example, the Capability Maturity Model Integration (CMMI) [10] addresses distinct AI-related concerns, especially ethical aspects, along with data integrity and model coordination requirements [11, 12]. AI frameworks can assist organizations in systematically discovering operational skill deficit areas while collaborating with established maturity models to guide technology investments and specify the deployment of suitable solutions based on the organization's current readiness levels [13]. Organizations with essential maturity struggle with data silos and knowledge gaps [12]. However, AI solutions (e.g., TensorFlow or PyTorch) need to be implemented strategically [14] as well as workforce preparation. Better matured organizations can unite artificial intelligence along with sophisticated strategies like generative AI and independent systems for complicated decision making and innovation [12]. The integration also helps to improve governance and compliance [15]. Explainability features and fairness mechanisms (e.g., AI Fairness 360 or LIME) that AI frameworks use to enforce some levels of rigor before products are put into production also benefit organizations through maturity models that help them inject such known-good criteria needed to satisfy regulators and society [12]. AI implementation achieves a better return on investment when the investment connections are directly tied to measurable organizational business results. McKinsey conducted broad-scale research [16] indicating that applying both AI structures and maturity models allows enterprises to deploy AI initiatives quickly and even quicker than with standalone AI tools. Results show that organizations following this combination are achieving 30% higher efficiency in adoption of AI projects than their strategic alignment. The intersection of AI frameworks and maturity models presents organizations with a system that not only progresses transformational AI adoption but also integrates sustainability and scalability into the journey. Hence, integrating AI frameworks with organizational readiness assessments are advancing AI implementations by building ethical ways of attaining meaningful outcomes across various sectors.

However, there is limited research on applying AI-driven approaches to cybersecurity maturity models for secure software coding. This research aims to address these gaps by introducing Generative AI (GenAI) approaches to enhance cybersecurity mitigation in secure software coding.

Here is how the rest of the paper is organized: Section 2 summarizes the research approach. Section 3 examines the research outcomes concerning our research inquiries. Section 4 is devoted to discussing the challenges to validity. Section 5 presents the conclusion and future research work.

## 2 Research Methodology

This study aimed to produce a holistic view of the scientific literature on secure software coding to understand the cybersecurity risks and practices available to address these risks. Established guidelines were implemented using a systematic literature review (SLR) [17-20]. The research question formulated in this study:

**RQ1:** What are the best GenAI practices to address cybersecurity risks to software development organizations in secure software coding?

According to the guidelines [17, 21-23] only the population and intervention variables are used to perform the systematic study. SLR (see Figure 1) is a preliminary study giving an overview of the status of research in a field [19, 24, 25]. Based on the population, intervention, and research question, we developed the following research string:
(("Software Coding" OR "Secure Software Development" OR "Software Implementation" AND ("Cybersecurity Risks" OR "Cybersecurity Issues" OR "GenAI Practices" OR "Generative Artificial Intelligence Models" OR "Artificial Intelligence Tools" OR "Artificial Intelligence Frameworks" OR "Artificial Intelligence Approaches" OR "Artificial Intelligence Techniques" OR "Artificial Intelligence Standards"))

We applied the search queries across the mentioned databases, examining all available fields. Table 1 shows the number of studies retrieved from each source. The collected papers were manually filtered based on the inclusion and exclusion criteria outlined in Table 2.

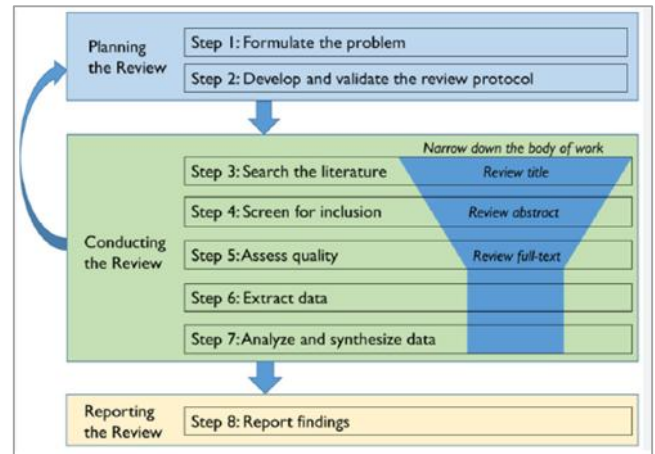

**Figure 1: Stage-Steps of Systematic Literature Review**

**Table 1: Number of Studies obtained in each Database**

| Digital Libraries | Search String Findings | Initial Selection | Final Selection |
|---|---|---|---|
| IEEE Xplore | 98 | 28 | 8 |
| ScienceDirect | 150 | 33 | 8 |
| ACM | 135 | 39 | 6 |
| SpringerLink | 230 | 50 | 4 |
| Google Scholar | 2021 | 60 | 5 |
| Total | 2694 | 220 | 31 |

**Table 2: Study Selection Criteria**

| Study Selection Criteria | |
|---|---|
| Inclusion Criteria (IC) | IC1: Articles must be published between the years 2010 and 2025 |
| | IC2: The article must be peer-reviewed and published in a Journal, conference, or workshop |
| | IC3: The article must be accessible online |
| | IC4: Article discussing some AI approaches, models, algorithms, considerations, challenges, risks, practices, and frameworks for cybersecurity in software coding |
| | IC5: Articles must be in English |

| Exclusion Criteria (EC) | EC1: Article not peer-reviewed |
|---|---|
| | EC2: The article is not a full research paper |
| | EC3: Article not in English |
| | EC4: Articles proposing AI approach, model, algorithm, consideration, challenge, risk, practice, and framework for cybersecurity in software coding |
| | EC5: The non-reputable publisher's articles |

Petersen et al. [21] suggest performing a quality assessment without setting strict criteria to avoid excluding potentially relevant systematic review resources. Therefore, we devise the following quality assessment approach (Table 3).

**Table 3: Study Quality Assessment Criteria**

| Quality Assessment Criteria (QAC) |
|---|
| QAC1: Relevance of the research problem |
| QAC2: Literature Review Quality |
| QAC3: Methodological Rigor |
| QAC4: Clarity and Accuracy of the Results |
| QAC5: Contribution to the Field |
| QAC6: Presentation and Writing Quality |

Following a comprehensive review of each primary study, we gathered data directly from the paper. This included qualitative and quantitative information: qualitative data comprised (i) GenAI practices for addressing cybersecurity risks. Thus, by outlining a clear map of the development of GenAI security practices in software coding, the paper offers an extensive study of the current trends and achievements in the field as well as the directions that require further research to prevent potential barriers to the successful implementation of secure software coding practices techniques.

The detailed report of the SLR results is prepared and structured under Section 3 of this paper. Therefore, through these simple and structured steps, the literature review will contribute to a comprehensive identification of the corresponding GenAI mitigation methods in secure software coding.

## 3 Results

Generative AI (GenAI) significantly contributes to detecting and preventing cybersecurity threats in secure software coding by leveraging its ability to analyze vast amounts of data, identify patterns, and predict potential vulnerabilities. A detailed exploration of its contributions, supported by a literature review, are presented in Table 4.

**Table 4: GenAI Practices for Addressing Cybersecurity Risks in Software Coding [5, 11, 26-38]**

| S. No | GenAI Practices for Insecure Coding Practices |
|---|---|
| 1 | Automated Code Analysis Tools |
| 2 | Secure Coding Standards Enforcements |
| 3 | Threat Modeling and Risk Assessment |
| 4 | AI-Powered Code Reviews |
| 5 | Secure Framework Recommendations |
| 6 | Training and Education |
| 7 | Monitoring and Logging Automation |
| 8 | CI/CD Pipeline Security Integration |
| 9 | Real-Time Vulnerability Patching |
| 10 | Phishing and Social Engineering AI |
| | **GenAI Practices for Vulnerable Dependencies** |
| 11 | Automated Dependency Scanning |
| 12 | Version Control and Dependency Updates |
| 13 | Risk Prioritization using AI Models |
| 14 | Natural Language Processing for Vulnerability Analysis |
| 15 | Machine Learning for Vulnerability Prediction |
| 16 | Dependency Mapping and Visualization |
| 17 | Secure Coding Guidelines Recommendation |
| 18 | Continuous Integration and Continuous Deployment (CI/CD) |
| 19 | Dynamic Vulnerability Testing |
| 20 | AI-Enhanced Software Composition Analysis (SCA) |
| | **GenAI Practices for Poor Error Handling** |
| 21 | Error Classification Models |
| 22 | Automated Vulnerability Scanning |
| 23 | Dynamic Runtime Monitoring |
| 24 | Predictive Code Analysis |
| 25 | Natural Language Processing (NLP) Bots |
| 26 | AI-Driven Test Case Generation |
| 27 | Automated Code Review |
| 28 | Adaptive Learning Systems |
| 29 | Self-Healing Code Frameworks |
| 30 | Error Handling Standards Enforcement |
| | **GenAI Practices for Weak Authentication/Authorization** |
| 31 | Enforce Strong Password Policies |
| 32 | Implement Multi-Factor Authentication |
| 33 | Use Secure Token-Based Authentication |
| 34 | Implement Session Management Controls |
| 35 | Adopt Secure Communication Protocols |
| 36 | Regular Security Testing |
| 37 | Centralized Identify Management |
| 38 | Monitor and Log Authentication Events |
| 39 | Continuous Training for Developers |
| | **GenAI Practices for Misconfigured Security Controls** |
| 40 | Automated Security Scanning |
| 41 | Real-Time Code Analysis |
| 42 | Intelligent Configuration Templates |
| 43 | Anomaly Detection in Code Behavior |
| 44 | Continuous Learning and Updates |
| 45 | Automated Policy Enforcement |
| 46 | Secure Coding Training via AI Simulation |
| 47 | Context-Aware Security Guidance |
| 48 | Vulnerability Prediction Models |
| | **GenAI Practices for Inadequate Encryption** |
| 49 | Use of Strong Encryption Algorithms |
| 50 | Secure Key Management |
| 51 | Implementation of AI-Powered Code Review |
| 52 | Regular Security Audits and Penetration Testing |
| 53 | Automated Vulnerability Scanning Tools |
| 54 | Developer Training in Cryptography |
| 55 | Secure Development Lifecycle (SDLC) Integration |
| 56 | Threat Modeling for Encryption Risks |
| 57 | AI-Driven Monitoring and Alerts |
| | **GenAI Practices for Cross-Site Scripting (XSS)** |
| 58 | Output Encoding |
| 59 | Context-Aware Escaping |
| 60 | Content Security Policy (CSP) |
| 61 | Avoid Inline JavaScript |
| 62 | Sanitize HTML |
| 63 | Educating Developers |
| 64 | Automated Testing Tools |
| 65 | Regular Code Review |
| 66 | Secure Defaults |
| **GenAI Practices for Insufficient Logging and Monitoring** | |
| 67 | Automated Log Analysis |
| 68 | Contextual Logging |
| 69 | Real-Time Monitoring with AI |
| 70 | Dynamic Log Level Adjustment |
| 71 | Predictive Maintenance through Log Insights |
| 72 | Correction Across Multiple Systems |
| 73 | Log Enrichment and Noise Reduction |
| 74 | Incident Root Cause Analysis |
| 75 | Data Privacy and Security in Logs |
| 76 | Feedback Loop for Log Optimization |
| **GenAI Practices for Race Conditions** | |
| 77 | Thread Synchronization |
| 78 | Avoid Shared State |
| 79 | Race Condition Testing |
| 80 | Atomic Operations |
| 81 | Thread-Safe Libraries |
| 82 | Deadlock Detection and Prevention |
| 83 | Code Reviews Focused on Concurrency |
| 84 | Formal Verification |
| 85 | Automated Tools for Race Condition Detection |
| 86 | AI-Assisted Static Code Analysis |
| 87 | Continuous Monitoring and Feedback Loops |
| 88 | Concurrency Design Pattern |
| 89 | Developer Training on Concurrency |
| **GenAI Practices for Inadequate Security Testing** | |
| 90 | Automated Vulnerability Scanning |
| 91 | AI-Driven Static Code Analysis |
| 92 | Dynamic Security Testing with AI |
| 93 | Predictive Threat Modeling |
| 94 | Anomaly Detection |
| 95 | Continuous AI-Base Monitoring |
| 96 | AI-Powered Penetration Testing |
| 97 | Adaptive Security Training |
| 98 | Secure Development Lifecycle Integration |
| 99 | Natural Language Processing (NLP) for Code Review |
| **GenAI Practices for Supply Chain Attacks** | |
| 100 | AI-Powered Dependency Analysis |
| 101 | Automated Risk Scoring for Components |
| 102 | Real-Time Supply Chain Monitoring |
| 103 | AI-Driven Threat Intelligence |
| 104 | Behavioral Analysis of Components |
| 105 | Automated Patch Management |
| 106 | Provenance Verification |
| 107 | Threat Simulation and Resilience Testing |
| 108 | Blockchain and AI Integration |
| 109 | Policy Enforcement with AI |

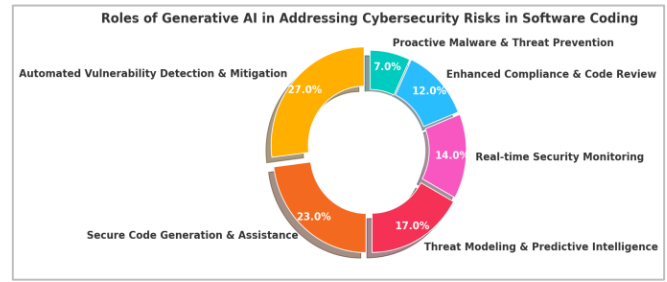

**Figure 2: Roles of GenAI in Addressing Cybersecurity Risks in Software Coding**

Figure 2 shows the role of GenAI in dealing with risks in cybersecurity towards software coding with six application areas.

Automated Vulnerability Detection and Mitigation accounts for the largest proportion of 27%. GenAI is very helpful to developers by automatically figuring out where the potential vulnerabilities are in code early on in the development cycle. It assists teams to find security weaknesses ahead of time and address them before they are exploited.

23% are Secure Code Generation and Assistance. This has shown GenAI capability of producing secure and robust code snippets or templates, which developers can use as it is in their applications reducing the amount of security flaws introduced through human error and improving the overall integrity of the software.

Threat Modeling and Predictive Intelligence accounts for 17 percent, identifying the segment that AI can also be used for potential security threat anticipation and security threat modeling. It utilizes historical data and current data regarding security trends to predict new threats, so that the developers and cybersecurity professionals can run more efficient defensive strategies.

Real time Security monitoring that represents 14% of how AI can monitor software systems effectively for anomaly and unusual activities. This continuous oversight enables rapid detection and following response to threats, reducing the potential losses from the cyberattacks.

12% Enhanced Compliance and Code Review suggests that Generative AI can help the developers to perform excessive compliance checks and all round code reviews. With it, teams can make sure that software is always operated according to the security standards and complies with the regulations of the industry, thereby minimizing the risk of compliance.

Finally, Proactive Malware and Threat Prevention is the last with 7%. Although the smallest part, this is an important area Generative AI is equipped with the capability to detect malicious patterns and threats early on before they can infect systems. The tools give developers and cybersecurity teams what they need to protect themselves against malware infections or cyber intrusion in their applications.

Together, these roles exemplify how Generative AI is revolutionizing the creation of secure, resilient, and trustworthy software environments and how it is becoming increasingly relevant in cybersecurity.

## 4 Limitations of the Study

This research is the first of its kind to provide a new systematic approach for AI-based cybersecurity protection in software development using GenAI practices, certain limitations need to be mentioned for better transparency and balanced discussion of the applicability, effectiveness as well as scalability of the presented model.

External validity refers to the generalizability of the findings to real-world contexts. The research foundation consists of conducting a systematic literature review (SLR) which examines existing studies about generative AI together with secure software practices. Multiple papers contained in the review provide broad coverage of software domains with AI implementations thus making the study results effectively transferable. The general application scope for particular industries together with emerging technologies demands additional specific studies beyond this research.

The research maintains internal validity through a systematic method of paper selection that uses transparent criteria for both inclusion and exclusion purposes. A standardized scoring system applied to research quality assessment leads to the selection of strong studies that reduce potential bias. The review controls potential confounders including study methodology differences through careful evaluation and discussion of quality assessments.

The review establishes construct validity by defining essential terms including "generative AI" alongside "secure software coding practices" and their clear relationship between these terms in this review. The chosen studies clearly explain their constructs while selecting papers dedicated to analyzing these constructs directly. The framework for categorization relies on established frameworks, which support consistency and relevance of analysis between practices and measures.

The study upholds high validity standards through strict methodological adherence to several validity dimensions, which yields dependable information about generative AI's role in developing secure software.

## 5 Conclusion and Future Research Direction

This study improves how well software companies maintain security during software development procedures by integrating GenAI practices. GenAI has proven effective in identifying vulnerabilities early in the software development lifecycle, preventing threats before they reach the final product. The SLR tracks security risks, detecting code injection vulnerabilities, securing against XSS attacks and preventing real-time buffer overflow incidents. The investigation shows why security should become a fundamental part of every stage of software development rather than an added-on element. Early implementation of cybersecurity strategies enables us to produce better-secured software systems while saving time and money from fixing vulnerabilities after deployment.

Future research should focus on:
- Integration with other AI Models: Future research could investigate opportunities to integrate ANN with AI paradigms, including but not limited to reinforcement learning and genetic algorithms, to further develop cybersecurity capabilities.
- Cross-Domain Applications: The ANN-ISM paradigm could be added to other domains such as cloud computing security, IoT security and blockchain-based cybersecurity frameworks.
- Implementation and Validation in Practice: Future work needs to adapt and evaluate this study in practice, integrated into the workflows of software development.

Our findings have significant implications for higher education and industry professionals by introducing a novel AI-driven approach to secure software development. The GenAI practices enhances cybersecurity by automating threat detection, reducing unauthorized access risks, and improving vulnerability management. Its cross-domain applicability extends to cloud security, IoT, and mobile application development, contributing to secure coding practices while improving cost efficiency and scalability. Moreover, its adoption could influence cybersecurity policies and regulatory frameworks. While the findings illustrate the potential of GenAI technology in enhancing cybersecurity, limitations such as data availability and quality, model complexity, generalization to diverse software environments, performance overhead, adaptability to evolving threats, and validation/testing challenges must be acknowledged. Future research should focus on enhancing the framework's adaptability to various security contexts and expanding its applicability across different scenarios to further strengthen software defense in our increasingly digital world.

## REFERENCES

[1] A. Alsirhani, M. Mujib Alshahrani, A. M. Hassan, A. I. Taloba, R. M. Abd El-Aziz, and A. H. Samak, "Implementation of African vulture optimization algorithm based on deep learning for cybersecurity intrusion detection," *Alexandria Engineering Journal,* vol. 79, pp. 105-115, 2023/09/15/ 2023.

[2] R. C. Chanda, A. Vafaei-Zadeh, H. Hanifah, and D. Nikbin, "Assessing cybersecurity awareness among bank employees: A multi-stage analytical approach using PLS-SEM, ANN, and fsQCA in a developing country context," *Computers & Security,* vol. 149, p. 104208, 2025/02/01/ 2025.

[3] A. Alzahrani and R. A. Khan, "Secure software design evaluation and decision making model for ubiquitous computing: A two-stage ANN-Fuzzy AHP approach," *Computers in Human Behavior,* p. 108109, 2023/12/26/ 2023.

[4] M. Humayun, M. Niazi, M. F. Almufareh, N. Z. Jhanji, S. Mahmood, and M. Alshayeb, "Software-as-a-Service Security Challenges and Best Practices: A Multivocal Literature Review," *Applied Sciences,* vol. 12, p. 3953, 2022.

[5] A. Ding, G. Li, X. Yi, X. Lin, J. Li, and C. Zhang, "Generative Artificial Intelligence for Software Security Analysis: Fundamentals, Applications, and Challenges," *IEEE Software,* pp. 1-8, 2024.

[6] G. Hanssen, C. Thieme, A. Bjarkø, M. Lundteigen, K. Bernsmed, and M. Jaatun, *A Continuous OT Cybersecurity Risk Analysis and Mitigation Process*, 2023.

[7] M. A. Akbar, K. Smolander, S. Mahmood, and A. Alsanad, "Toward successful DevSecOps in software development organizations: A decision-making framework," *Information and Software Technology,* vol. 147, p. 106894, 2022/07/01/ 2022.

[8] M. A. Akbar, S. Rafi, S. Hyrynsalmi, and A. A. Khan, "Towards People Maturity for Secure Development and Operations: A vision," presented at the Proceedings of the 28th International Conference on Evaluation and Assessment in Software Engineering, Salerno, Italy, 2024.

[9] R. Ulfsnes, N. Moe, V. Stray, and M. Skarpen, *Transforming Software Development with Generative AI: Empirical Insights on Collaboration and Workflow*, 2024.

[10] S. U. Team, "Standard CMMI Appraisal Method for Process Improvement (SCAMPI) A, Version 1.3: Method Definition Document," HANDBOOK

CMU/SEI-2011-HB-001March 2011.

[11] A. Gurtu and D. Lim, "Chapter 101 - Use of Artificial Intelligence (AI) in Cybersecurity," in *Computer and Information Security Handbook (Fourth Edition)*, J. R. Vacca, Ed., ed: Morgan Kaufmann, 2025, pp. 1617-1624.

[12] A. O. Almagrabi and R. A. Khan, "Optimizing Secure AI Lifecycle Model Management with Innovative Generative AI Strategies," *IEEE Access,* pp. 1-1, 2024.

[13] M. Islam, F. Khan, S. Alam, and M. Hasan, *Artificial Intelligence in Software Testing: A Systematic Review*, 2023.

[14] I. Jada and T. O. Mayayise, "The impact of artificial intelligence on organisational cyber security: An outcome of a systematic literature review," *Data and Information Management,* vol. 8, p. 100063, 2024/06/01/ 2024.

[15] M. Ilyas, S. U. Khan, H. U. Khan, and N. Rashid, "Software integration model: An assessment tool for global software development vendors," *Journal of Software: Evolution and Process,* vol. n/a, p. e2540, 2023.

[16] M. Quarterly, "Implementing generative AI with speed and safety," https://www.mckinsey.com/capabilities/risk-and-resilience/our-insights/implementing-generative-ai-with-speed-and-safety, pp. 1-10, 2024.

[17] B. Kitchenham and S. Charters, "Guidelines for Performing Systematic Literature Reviews in Software Engineering," Keele University and Durham University2007.

[18] R. A. Khan, S. U. Khan, H. U. Khan, and M. Ilyas, "Systematic Literature Review on Security Risks and its Practices in Secure Software Development," *IEEE Access,* vol. 10, pp. 5456-5481, 2022.

[19] B. Kitchenham, O. Pearl Brereton, D. Budgen, M. Turner, J. Bailey, and S. Linkman, "Systematic literature reviews in software engineering – A systematic literature review," *Information and Software Technology,* vol. 51, pp. 7-15, 2009/01/01/ 2009.

[20] R. A. Khan, S. U. Khan, M. Alzahrani, and M. Ilyas, "Security Assurance Model of Software Development for Global Software Development Vendors," *IEEE Access,* vol. 10, pp. 58458-58487, 2022.

[21] K. Petersen, S. Vakkalanka, and L. Kuzniarz, "Guidelines for conducting systematic mapping studies in software engineering: An update," *Information Software Technology,* vol. 64, pp. 1-18, 2015.

[22] K. Petersen, R. Feldt, S. Mujtaba, and M. Mattsson, "Systematic mapping studies in software engineering," *Proceedings of the 12th International Conference on Evaluation and Assessment in Software Engineering,* pp. 68–77, 2008.

[23] R. A. Khan, S. A. Ullah, and I. Yazid, "Systematic mapping study protocol for secure software engineering," in *Proc. Asia Int. Multidisciplinary Conf.(AIMC)*, 2019, pp. 367-374.

[24] A. R. Khan, S. Khan, and M. Ilyas, *Exploring Security Procedures in Secure Software Engineering: A Systematic Mapping Study*, 2022.

[25] R. A. Khan, S. U. Khan, M. A. Akbar, and M. Alzahrani, "Security risks of global software development life cycle: Industry practitioner's perspective," *Journal of Software: Evolution and Process,* vol. 36, pp. 1-34, 2023.

[26] A. O. Almagrabi and R. A. Khan, "Optimizing Secure AI Lifecycle Model Management With Innovative Generative AI Strategies," *IEEE Access,* vol. 13, pp. 12889-12920, 2025.

[27] F. Alwahedi, A. Aldhaheri, M. A. Ferrag, A. Battah, and N. Tihanyi, "Machine learning techniques for IoT security: Current research and future vision with generative AI and large language models," *Internet of Things and Cyber-Physical Systems,* vol. 4, pp. 167-185, 2024/01/01/ 2024.

[28] Y. Bajaj and M. Samal, "Accelerating Software Quality: Unleashing the Power of Generative AI for Automated Test-Case Generation and Bug Identification," *International Journal for Research in Applied Science and Engineering Technology,* 2023.

[29] Z. Cai, Z. Xiong, H. Xu, P. Wang, W. Li, and Y. Pan, "Generative Adversarial Networks: A Survey Toward Private and Secure Applications," *ACM Comput. Surv.,* vol. 54, p. Article 132, 2021.

[30] D. Dalalah and O. M. Dalalah, "The false positives and false negatives of generative AI detection tools in education and academic research: The case of ChatGPT," *The International Journal of Management Education,* vol. 21, p. 100822, 2023.

[31] C. Ebert and P. Louridas, "Generative AI for software practitioners," *IEEE Software,* vol. 40, pp. 30-38, 2023.

[32] M. A. Ferrag, F. Alwahedi, A. A. Battah, B. Cherif, A. Mechri, N. Tihanyi*, et al.*, "Generative AI in Cybersecurity: A Comprehensive Review of LLM Applications and Vulnerabilities," 2024.

[33] R. Gupta and B. Rathore, "Exploring the generative AI adoption in service industry: A mixed-method analysis," *Journal of Retailing and Consumer Services,* vol. 81, p. 103997, 2024/11/01/ 2024.

[34] F. P. Levantino, "Generative and AI-powered oracles: "What will they say about you?"," *Computer Law & Security Review,* vol. 51, p. 105898, 2023/11/01/ 2023.

[35] T. McIntosh, T. Liu, T. Susnjak, H. Alavizadeh, A. Ng, R. Nowrozy*, et al.*, "Harnessing GPT-4 for generation of cybersecurity GRC policies: A focus on ransomware attack mitigation," *Computers & Security,* vol. 134, p. 103424, 2023/11/01/ 2023.

[36] G. S. Nadella, S. R. Addula, A. R. Yadulla, G. S. Sajja, M. Meesala, M. H. Maturi*, et al.*, "Generative AI-Enhanced Cybersecurity Framework for Enterprise Data Privacy Management," *Computers,* vol. 14, p. 55, 2025.

[37] C. Novelli, F. Casolari, P. Hacker, G. Spedicato, and L. Floridi, "Generative AI in EU law: Liability, privacy, intellectual property, and cybersecurity," *Computer Law & Security Review,* p. 106066, 2024/10/25/ 2024.

[38] S. Sai, U. Yashvardhan, V. Chamola, and B. Sikdar, "Generative AI for Cyber Security: Analyzing the Potential of ChatGPT, DALL-E and Other Models for Enhancing the Security Space," *IEEE Access,* vol. PP, pp. 1-1, 01/01 2024.